\documentclass[a4paper]{article}
\usepackage{nameref,hyperref}
\usepackage[utf8]{inputenc}
\usepackage[english]{babel}
\usepackage{amsmath}
\usepackage{graphicx}
\usepackage{multirow}
\usepackage[colorinlistoftodos]{todonotes}

\title{Simulating the interaction of road users: A glance to complexity of Venezuelan traffic}

\author{Juan C. Correa$^{1\footnote{This author finished this work while being at Universidad Simón Bolívar in Caracas - Venezuela. His permanent affiliation is at Fundación Universitaria Konrad Lorenz in Bogotá - Colombia. E-mail: \url{juanc.correan@konradlorenz.edu.co}}}$, Mario I. Caicedo$^{3}$,\\ Ana L. C. Bazzan$^{2}$ \& Klaus Jaffe$^{4}$}

\date{\small$^{1}$ Facultad de Psicología. Fundación Universitaria Konrad Lorenz. Colombia\\
$^{2}$ 
Instituto de Informática. Universidade Federal do Rio Grande do Sul. Brazil\\
$^{3}$
Departamento de Física. Universidad Simón Bolívar. Venezuela\\
$^{4}$
Centro de Estudios Estratégicos. Universidad Simón Bolívar. Venezuela}

\begin{document}
\maketitle

\begin{abstract}

Automotive traffic is a classical example of a complex system, being the simplest case the homogeneous traffic where all vehicles are of the same kind, and using different means of transportation increases complexity due to different driving rules and interactions between each vehicle type. In particular, when motorcyclists drive in between the lanes of stopped or slow-moving vehicles. This later driving mode is a Venezuelan pervasive practice of mobilization that clearly jeopardizes road safety. Here, we developed a minimalist agent-based model to analyze the interaction of road users with and without motorcyclists on the way. The presence of motorcyclists dwindles significantly the frequency of lane changes of motorists while increasing their frequency of acceleration/deceleration maneuvers, without significantly affecting their average speed. That is, motorcyclist ``corralled" motorists in their lanes limiting their ability to maneuver and increasing their acceleration noise.  Comparison of the simulations with real traffic videos shows good agreement between model and observation. The  implications of these results regarding road safety concerns about the interaction between motorists and motorcyclists are discussed.
\end{abstract}

\section{Introduction}

\begin{figure}[h!]
\centering
\includegraphics[width=0.3\textwidth]{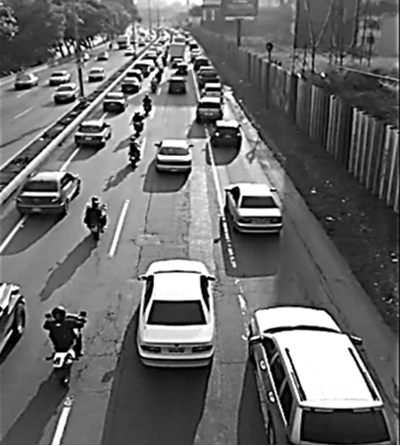}
\caption{\label{fig1}The practice of motorcycle lane-sharing in heterogeneous traffic}
\end{figure}

An almost worldwide practice of mobilization in urban traffic is known as ``motorcycle lane-sharing''. This practice occurs when motorcyclists ride in between the lanes of stopped or slow-moving vehicles; an event observed in the so-called “heterogeneous traffic” in which different types of vehicles interact on the road (see Figure \ref{fig1}).

% {\bf{¿de veras crees que los dos parrafitos que siguen son relevantes para nosotros?, Es algo psicológico y no tenemos soporte alguno para discutir nada de esa índole}} \color{blue}{Lo voy a quitar}

% Notwithstanding that motorcyclists acknowledge the factors that increase the likelihood of accidents \cite{Mannering1995} it is unclear if these events are due to other vehicles violating their right of way \cite{Ohlhauser2011} or because of the “aggressiveness” of motorcyclists in riding on the road \cite{Jones2013} or an interaction of these factors. However, motorcycle lane-sharing also has psychological implications for road users \cite{Brooks1985}.

% In general, motorists  have negative attitudes towards motorcyclists and how they filter through traffic \cite{Crundall2008} . These attitudes might be due to the absence of specific traffic regulations \cite{Sperley2010}. For example, reports from India show a lack of “lane discipline” because individuals do not drive inside their designated lanes \cite{doi:10.3846/16484142.2014.928788}.

The practice of motorcycle lane-sharing has been studied with computer simulations. Lan and Chang \cite{Lan2003} simulated the moving behaviors of motorbikes in a mixed traffic with cars, and concluded that the maximum flow rates corresponded to a relative road occupancy (a proxy of density) of 17\% to 30\%. In another study, Lan and Chang \cite{Lan2005} observed that the speed of motorcyclists was slightly higher than the motorists and that both maximum flow rates and critical speeds declined with an increase of maximum speed deviations. Lee \cite{Lee2007} modeled six behaviors of motorcyclists (i.e., traveling alongside another vehicle, oblique following, filtering, moving to the head of a queue, swerving or weaving, and tailgating) and concluded that the presence of motorcycles enlarged the capacity of the road; b) the installation of a motorcycle’s dedicated lane on a road increased the maximum flow rate by around 20\% because of the additional
capacity of the motorcycle lane; c) the values of the “passenger car unit” (a metric used to assess heterogeneous traffic-flow rate) of motorcycles were considerably higher than the values of the passenger car unit of four-wheeled vehicles, since motorbikes are still able to progress by filtering when the movements of cars are constrained by density. Lan and his colleagues \cite{Lan2010} simulated the interaction between motorists and motorcyclists and observed a larger deterioration of car flow as the number of motorcycles increased.

% {\color{red} {Some simulations focused on understanding the explicit rules that govern the individual driving behavior; with the “car-following models” \cite{Brackstone1999,Brackstone2009} and the “lane-change models” \cite{Hidas2002,Hidas2005} serving as conceptual frameworks \cite{Toledo2007}. In Asia, these developments included the application of models, in urban and rural settings with several road geometries like highways, ramps or intersections \cite{Krishnamurthy2014}.  In Europe some researchers aimed to include the role of psychological factors such as “politeness” in letting others change lane \cite{Kesting2007}. Many examples of traffic simulations can be found in the literature \cite{doi:10.3846/16484142.2014.928788,Mallikarjuna2011}, but experimental validation of simulations are rare \cite{Brockfeld2006}. The analysis of individual trajectories of vehicles and motorcycles allows developing “microscopic models” of heterogeneous traffic \cite{Chunchu2010}. Such procedures produce quantitative mismatches between the model and real-life data that are normally in the range between 11\% and 29\% for variables such as density, speed and flow \cite{Kesting2008a}.}}

\begin{figure}[h!]
\centering
\includegraphics[width=1\textwidth]{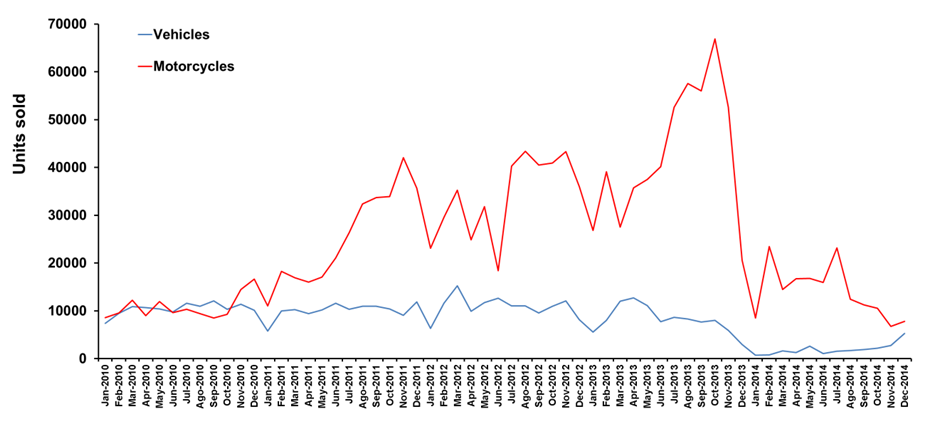}
\caption{\label{fig2}Monthly sales of motorcycles and cars in Venezuela.}
\end{figure}

In Venezuela the use of motorcycles has dramatically increased as reflected in the rising sales of motorcycles during the last five years although dropping lately because of an acute economic and political crisis (see Figure \ref{fig2}). As a consequence, the practice of motorcycle lane-sharing has proliferated increasing road safety concerns \cite{Wigan2002}. Only in the first ten months of 2012 more than three daily fatalities of motorcyclists in road accidents were reported \cite{Figuera2012}.

Our purpose in this work is to address the case of Venezuelan motorcyclists, who often ride in between other vehicles. We are interested in capturing the effect of such behavior on motorists’ behavior, if any. We tackle this problem through a minimalist agent-based model capturing the most basic interactions between road users. Agents' rules were devised in order to mimic what any driver would observe when using a Venezuelan highway, i.e. extremely aggressive motorcyclists not afraid of driving between lanes at roughly $40$ mph. The rest of the paper is organized as follows. Section 2 presents the procedure and results of empirical observations conducted in a Venezuelan highway. Section 3 describes the agent-based model. Section 4 presents the results of our simulation model and section 5 discusses the implications of our overall results.

 \section{Caracas traffic as it is}

The reader can watch one of our observations in
youtube (\url{http://youtu.be/QlCIqaXOQzY}). This video shows the segment of “Francisco Fajardo Highway” in the south-west of Caracas where we observed the two types of traffic: homogeneous traffic where only motorists were on the road and heterogeneous traffic where motorcyclists and motorists used the road at the same time. Motorcyclists ride in between the left and the central lane of the road, while homogeneous traffic occurs between the central and the right lane where a motorcycle was rarely seen. The videos were recorded when traffic was high (7:00am - 8:00am and 17:00pm - 18:00pm) during working days.

The analysis of heterogeneous traffic flow required the use of the so-called passenger
car unit (“PCU”) which is a proxy of density based on the idea of the physical occupation of the road that corresponds to each type of vehicle. The PCU values of cars and motorcycles have been empirically determined \cite{Lee2007}; a 1 PCU represents a standard four-wheeled vehicle and 0.4 PCUs represent a standard motorcycle. These values were employed to estimate the heterogeneous traffic flow of our videos and found that this was on average 1.25 times higher than the homogeneous flow; indicating that the practice of motorcycle lane-sharing increases the physical capacity of the road by allowing a greater volume of heterogeneous traffic. The presence of motorcycles affected the frequency of lane-changing maneuvers and the speed of motorists. From the analysis of 20 videos we observed that motorists changed lanes 9.9 $\pm$ 6.6 times per minute (average $\pm$ standard deviation) in homogeneous traffic and 3.7 $\pm$ 2.9 times in heterogeneous traffic. The average speed of motorists proved to be quite similar in both situations (9.95 $\pm$ 2.6 m/s in homogeneous
traffic versus 9.9 $\pm$ 2.2 m/s in heterogeneous traffic). A Student's t-test showed that the
difference in lane changing frequency was statistically significant (t $=$ 18.05; df $=$ 1;
p $<$ 0.01), between both type of traffic, whereas that between the average speeds was not (t = 1.02; df = 1; p $>$ 0.1).

\section{A model of motorcycle lane-sharing}

The model was implemented with “SeSAm” and is available for download at (\url{http://www.openabm.org/model/3135/version/2/view}).
%, an open-source software designed for social scientists that has been used for traffic simulations before \cite{Klugl2004}. This software is available at the web page \url{http://simsesam.de/} and the model. 
The model simulates cars traveling on a two lane road with and without interacting with motorcyclists. The purpose of our model is to explore the effects of motorcycle lane-sharing on motorist's behavior in order to answer the following questions: 1) What is the quantitative impact of motorcycle lane-sharing on the motorist's behavior of lane-changing? 2) What is the quantitative impact of motorcycle lane-sharing on the frequency of acceleration and deceleration maneuvers performed by motorists? 3) What is the effect of motorcycle lane-sharing on the average speed of motorists? Our procedure of data collection includes the observation of average speed; the frequency of accelerations/deceleration maneuvers and the frequency of lane-changing maneuvers of motorists and motorcyclists. These metrics have been deemed relevant for traffic management and traffic safety \cite{Kesting2007,Laval2006}. Our agent-based model reproduces the behavioral interaction between road users and allowed for quantitative and for visual qualitative presentation of the results.

The model comprises three agents: the road, the motorists and the motorcyclists. These agents interact in a non-toroidal 2-dimensional virtual environment, represented as a map with a coordinate system whose origin lies in the upper, left corner. The physical dimensions of this map were 40 meters of height by 400 meters of width. The physical dimensions of the road were equivalent to a two lane avenue with a length of 400 meters. The width of each lane was fixed to 4 meters. The total length of a single lane is constituted by 100 cells of 4 meters length by 4 meters width. We assume a maximum speed of 10 m/s. These features are common to the so-called “cellular automata models” where the road is divided as a grid and each cell behaves according to standard rules that allow the simulation of the macroscopic behavior of traffic (see for instance \cite{Li2006}). The cells of the top lane of the road were identified with the number one and the cells of the bottom lane were identified with
the number two for allowing the motorists to change lanes according to the rules
described in Table \ref{tab1}.

\begin{table}[h!]
\centering
\caption{State variables of motorists and motorcyclists}
\label{tab1}
\begin{tabular}{|l|l|l|}
\hline
Name & \begin{tabular}[c]{@{}l@{}}Type of variable\\ and default values\end{tabular} & Description                                                                                                                                                                                \\ \hline
D    & \begin{tabular}[c]{@{}l@{}}Numeric\\ (Direction = 180)\end{tabular}           & \begin{tabular}[c]{@{}l@{}}Stands for direction \\ (from east to west)\end{tabular}                                                                                                        \\ \hline
IS   & \begin{tabular}[c]{@{}l@{}}Numeric\\ (5, 6, 7, 8, 8, 10)\end{tabular}         & \begin{tabular}[c]{@{}l@{}}Stands for instantaneous speed\\ in m/s\end{tabular}                                                                                                            \\ \hline
PS   & \begin{tabular}[c]{@{}l@{}}Numeric\\ (5, 6, 7, 8, 9, 10)\end{tabular}         & \begin{tabular}[c]{@{}l@{}}Stands for preferred speed\\ in m/s\end{tabular}                                                                                                                \\ \hline
CL   & \begin{tabular}[c]{@{}l@{}}Numeric\\ (1, 2)\end{tabular}                      & \begin{tabular}[c]{@{}l@{}}Identifies the lane in which the vehicle is \\ circulating at aninstantaneous moment\end{tabular}                                                               \\ \hline
P    & \begin{tabular}[c]{@{}l@{}}Position\\ (Cartesian coordinates)\end{tabular}    & \begin{tabular}[c]{@{}l@{}}Indicates the instantaneous position \\ of a driver inside the road.\end{tabular}                                                                               \\ \hline
IFV  & List of entities                                                              & \begin{tabular}[c]{@{}l@{}}Enumerates in no specific order\\ the list of in-front vehicles inside a\\ radius of 25 meters from the\\ instantaneous position of a vehicle.\end{tabular}     \\ \hline
FV   & List of entities                                                              & \begin{tabular}[c]{@{}l@{}}Enumerates in no specific order \\ the list of following vehicles inside a \\ radius of 25 meters from the \\ instantaneous position of a vehicle.\end{tabular} \\ \hline
VAL  & List of entities                                                              & \begin{tabular}[c]{@{}l@{}}Set that enumerates in no specific \\ order the list of vehicles that are seen \\ from the left rear view mirror\end{tabular}                                   \\ \hline
VAR  & List of entities                                                              & \begin{tabular}[c]{@{}l@{}}Set that enumerates in no specific \\ order the list of vehicles that are seen \\ from the right rear view mirror\end{tabular}                                  \\ \hline
\end{tabular}
\end{table}

The road allowed the entrance of cars on the right edge of the way, inside a randomly selected lane and with a randomly selected preferred speed. The critical reader might be against this form of modeling arguing that traffic flow does not behave completely at random. Yet, this form of modeling has been done in previous simulations \cite{Arasan2005,Krishnamurthy2014}, and it was not our purpose to study traffic flow transitions and its impact on mixed traffic. The physical
dimensions of cars were set to 4.5 meters length by 1.7 meters width, resembling a “standard compact vehicle” in the real world and motorcycles were set to 2 meters length by 1 meter width, resembling a “standard street motorbike”. Regarding the driving behavior of motorists, we used the geometrical central point of the cells of the road to locate the position of cars “inside the lane” when entering to the road. Motorcycles were located at the bottom of the first lane which corresponds to the space in between cars
(i.e., between lane 1 and lane 2). Motorcyclists could “switch lanes” but never traveled
“inside” them. Motorists and motorcyclists decided their instantaneous behavior following
the rules defined in the format of “if-then-else” decisions that are summarized in Table \ref{tab2}.
We are keenly aware of the simplistic rules that we use to mimic the current behavior of
motorists and motorcyclists. Yet, we justify this approach because it allows understand
how simple rules interact to create the macroscopic behavior of traffic which is
commonly analyzed in terms of macroscopic variables that do not consider the
behavioral interactions among road users.

\begin{table}[h!]
\centering
\caption{Behavioral rule for motorists and motorcyclists}
\label{tab2}
\begin{tabular}{|l|l|l|}
\hline
Name of the action                                                          & Check IF                                                                     & Then, do... (Else)                                                                                                                                       \\ \hline
Accelerate                                                                  & IS $<$ PS                                                                    & \begin{tabular}[c]{@{}l@{}}Change speed by 1 m/s\\ (Otherwise, if IS $>$ PS, \\ and change IS to PS)\end{tabular}                                               \\ \hline
Decelerate                                                                  & IFV $\geq$ 1                                                                 & \begin{tabular}[c]{@{}l@{}}Change speed by 4 m/s\\ (If IS $<$ 0, set IS = 0)\end{tabular}                                                                \\ \hline
\begin{tabular}[c]{@{}l@{}}Switch to \\ right lane\end{tabular}             & \begin{tabular}[c]{@{}l@{}}CL = 2\\ IFV $\geq$ 1\\ VAR = 0\end{tabular}      & Adds 45 degrees to the value of D                                                                                                                        \\ \hline
\begin{tabular}[c]{@{}l@{}}Enter into the \\ right target lane\end{tabular} & \begin{tabular}[c]{@{}l@{}}D = 225\\ CL = 2\end{tabular}                     & Set D = 180 and set CL = 1                                                                                                                               \\ \hline
\begin{tabular}[c]{@{}l@{}}Switch to \\ the left lane\end{tabular}          & \begin{tabular}[c]{@{}l@{}}CL = 1\\ IFV $\geq$ 1\\ VAL = 0\end{tabular}      & Subtracts 45 degrees to the value of D                                                                                                                   \\ \hline
\begin{tabular}[c]{@{}l@{}}Enter into the \\ left target lane\end{tabular}  & \begin{tabular}[c]{@{}l@{}}D = 135\\ CL = 1\end{tabular}                     & Set D to 180 and set CL to 2                                                                                                                             \\ \hline
Keep speed                                                                  & \begin{tabular}[c]{@{}l@{}}None of the \\ previous \\ rules met\end{tabular} & \begin{tabular}[c]{@{}l@{}}Move according to IS , \\ Set P and CL, \\ Set the list of moving entities \\ reflected in all rear view mirrors\end{tabular} \\ \hline
\end{tabular}
\end{table}
\newpage
The model proceeds in discrete time steps and the road maintains a constant vehicular
traffic flow by verifying the maximum number of cars that should be created for allowing
their controlled entrance in either lane at the right edge of the map, and eliminating them
when reaching the left edge of the road. This procedure is similar to previous simulations \cite{Arasan2010,Arasan2005} and this process iterates during 700 time steps for each simulation run. In the first 100 time steps no data is gathered in order to reach an invariant traffic flow (this is necessary for experimental control and it is not intended for mimicking “realistic” traffic flow transitions). Thus, data collection was done from t = 101 until t = 700. Once moving agents were created their state variables were
processed in a random order within each time step. The motorists-motorcyclists ratio was kept constant at 2:1.
%\section{Experimental design and simulation}
Our experiment follows a completely randomized factorial design \cite{Fisher1935}. We manipulated traffic flow by creating five discrete levels of density (4, 8, 12, 16 and 20
cars on the road) in two types of traffic (homogeneous and heterogeneous). Both motorists and motorcyclists were allowed to drive with different preferred speeds as described in Table \ref{tab1}. The experimental setup used different combinations of density and traffic type, resulting in ten experimental situations. In order to make a tractable experiment, we limited the preferred speed of motorists and motorcyclists to 10 m/s. Data was summarized in a new database containing the motorists’ performance for each time step in 24 replications for each experimental situation. 

\section{Results and validation of the model}

\begin{figure}[h]
\centering
\includegraphics[width=0.95\textwidth]{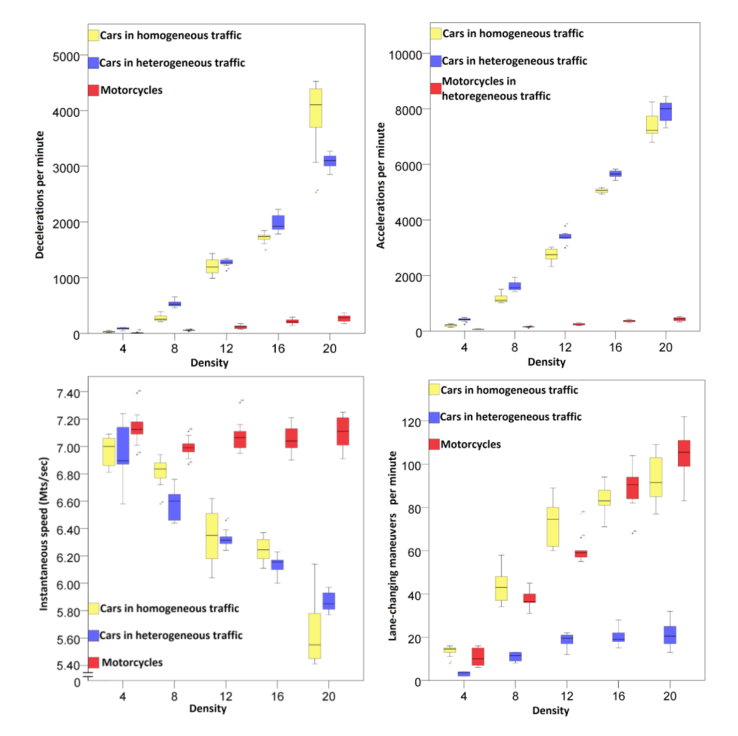}
\caption{\label{figs}Statistical distributions of driving maneuvers and average speed of road users as a function of traffic density and traffic type}
\end{figure}

The statistical distributions for the frequency of driving maneuvers as well as average speed of both motorists and motorcyclists are depicted in Figure \ref{figs}. Note that acceleration/decelerations maneuvers of motorists increased when density increased in  both traffic types, although they were performed less often in the homogeneous traffic condition  ($F_{acceleration} $= 1148.40; df = 4; p $<$ 0.001; $F_{deceleration} $= 527.45; df = 4; p $<$ 0.001). The 95.8\% of the variance for motorists’ decelerations, and the 98.1\% of the variance for motorists' accelerations were mainly accounted for by density ($\Omega^{2}_{acceleration}$ = 93.9\%; $\Omega^{2}_{deceleration}$ = 88\%) than by traffic type ($\Omega^{2}_{acceleration}$ = 0.7\%; $\Omega^{2}_{deceleration}$ = 1.1\%). The average speed of motorists dwindled while density increased, but motorcyclists kept their average speed regardless of density ($F_{speed} $= 258.44; df = 4; p $<$ 0.001) and almost all of its statistical variance was accounted for by density ($\Omega^{2}$ = 88.5\%). Interestingly enough, the frequency of lane-changing maneuvers  increased when density increased for both traffic types ($F_{lane-changing} $= 89.64; df = 4; p $<$ 0.001), but motorists changed lanes four times more in the homogeneous traffic condition compared with the heterogeneous traffic, and motorcyclists changed lanes as frequent as motorists in the heterogeneous traffic condition. The larger part of the
variance of lane-changing maneuvers was accounted for by the traffic type (F = 536.22; df = 1; p $<$ 0.001; $\Omega^{2}$ = 31.2\%), the density (F = 110.26; df = 4; p $<$ 0.001; $\Omega^{2}$ = 30.6\%) and the combination of these factors (F = 89.64; df = 4; p $<$ 0.001; $\Omega^{2}$ = 25.1\%).

\begin{figure}[h!]
\centering
\includegraphics[width=0.8\textwidth]{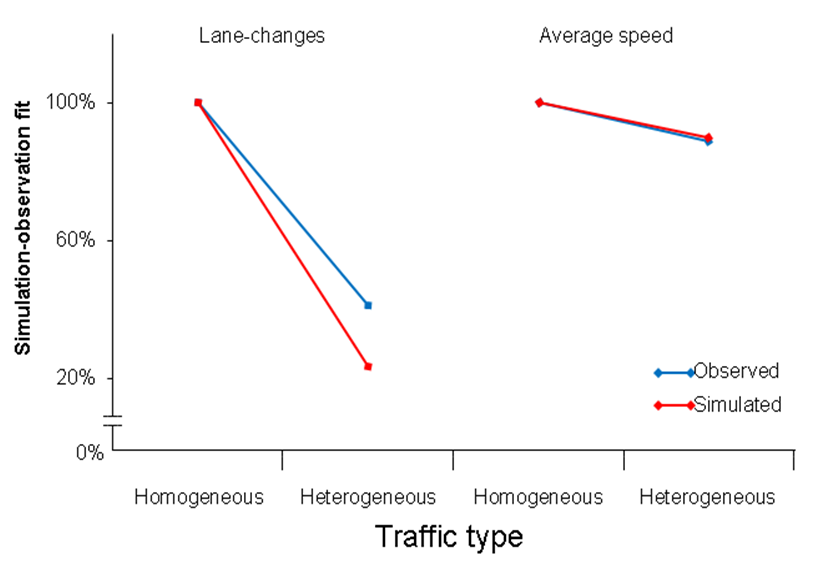}
\caption{\label{fig9}Comparison between simulated and real traffic data. Data for homogeneous
traffic is set to be 100\% in both cases.}
\end{figure}

The validity of these results were assessed through a measure of agreement for interval or nominal multivariate observations ``by different sets of judges” \cite{Janson2004}. This measure, known as “iota” ($\iota$), indicates an agreement when a set of two or more observers have rated a sample of objects on several dimensions or variables. This measure can be interpreted in terms of expected and observed disagreement,
calculated as average distances between judges’ observations, with a correction factor
of agreement by chance. To estimate the agreement between the results of our model and the results of our observations, we created a unified table composed by data for relative differences between traffic types of the frequency of lane-changing maneuvers performed by virtual and real motorists as well as the difference of their average
instantaneous speed, controlling for traffic density. This procedure is useful for validating
“ordinal patterns” \cite{Thorngate2013} between the two types of traffic. The agreement between the simulated results and the observed results proved to be highly significant ($\iota$ = 92,91\%), as depicted in Figure \ref{fig9}.

\newpage

\section{Discussion}
The effects of motorcycle lane-sharing on motorists’ behavior are consistent with Lee’s
earlier results showing that heterogeneous traffic flow is higher than the
homogeneous one \cite{Lee2007}. In addition, our results complement previous findings described by
Lan and colleagues \cite{Lan2010} on the incentives for motorists to make lateral drifts and lane changes when sharing the lane with motorcyclists, showing that motorcycle lane-sharing ``corrals'' motorists into their lane reducing their frequency of lane changes while
increases their frequency of acceleration/deceleration maneuvers. Also, the increase in acceleration and deceleration maneuvers increased acceleration noise (standard
deviation of acceleration), which in turn affects the level of service of the roadway, and is
also sometimes related to safety \cite{Boonsiripant2009}.

The above effects have to be combined with the fact that the average overall speed of
motorists proved to be much more affected by traffic density than by the presence of
motorcyclists on the road, which suggests that traffic accidents in heterogeneous traffic
are not necessarily due to motorists’ speed, but to the maneuvers of both drivers and
motorcyclists, which is consistent with previous simulations designed to analyze car
accidents in homogeneous traffic \cite{Moussa2003} and earlier experiments on the interaction between motorists and motorcyclists in experimental settings \cite{Ohlhauser2011}.

Our study is by no means exhaustive on the observed complexities of Venezuelan traffic. Yet, we regard our model as one that illustrates how very simplistic rules for driving behavior can mimic traffic dynamics, without including very sophisticated features of simulation. Future works could include more detailed aspects of (un)safe driving behavior, such as the analysis of traffic accidents that are due to driving maneuvers, speed and their combination. As a final remark, we believe that our results may be significant for the research that aims studying the link between physical parameters of traffic and human behavior.

\section{Acknowledgements}
This research used resources of ``Centro de Computo Científico del Grupo de Relatividad y Campos'' of Universidad Simón Bolívar, Caracas – Venezuela. We are also
indebted with Tess Roseng Carbonell for her assistance in reviewing preliminary
versions of this manuscript.

\bibliographystyle{ieeetr}
\bibliography{REFS.bib}
\end{document}